# Comment on "Stochastic local operations and classical communication invariant and the residual entanglement for *n* qubits"

Xin-Wei Zha[*], Hai-Yang Song, Ming-Liang Hu

Department of Applied Mathematics and Applied Physics, Xi'an Institute of Post and Telecommunications, Xi'an, 710121, P R China

**Abstract**: In a recent paper [Phys. Rev. A 76, 032304(2007)], Li et al. proposed the definition of the residual entanglement for *n* qubits by means of the stochastic local operations and classical communication (SLOCC). Here we argue that their definition is not suitable for the case of odd-*n* qubits.

Recently, Li et al. establish a relation between stochastic local operations and classical communication (SLOCC) and residual entanglement [1]. In their paper, they gave the definition of the residual entanglement for odd-*n* qubits by means of the invariant, which is defined as

$$\tau(\psi) = 4\left| \left[\bar{\mathcal{I}}(a,n)\right]^2 - 4\mathcal{I}^*(a,n-1)\mathcal{I}^*_{+2^{n-1}}(a,n-1) \right|. \tag{1}$$

where $[\bar{\mathcal{I}}(a,n)]^2 - 4\mathcal{I}^*(a,n-1)\mathcal{I}^*_{+2^{n-1}}(a,n-1)$ is the invariant for odd-*n* qubits, and

$$\tau(\psi) = \tau(\psi')\left|\det^2(\alpha)\det^2(\beta)\det^2(\gamma)\cdots\right| \tag{2}$$

Here, we discuss the weaknesses of this proposal.

For arbitrary odd-*n* qubit case, we can obtain the polynomial invariant [2]

$$Z^k_{12\cdots n} = \langle\psi|T_1 T_2 \cdots \sigma_{kx} \cdots T_{n-1}T_n|\psi^*\rangle^2 + \langle\psi|T_1 T_2 \cdots \sigma_{kz} \cdots T_{n-1}T_n|\psi^*\rangle^2 \\ - \langle\psi|T_1 \cdots T_{k-1}T_{k+1}\cdots T_{n-1}|\psi^*\rangle^2 \tag{3}$$

where $T_k = i\sigma_{ky}$. For the case of *n*=3 and 5, it s easy to show the polynomial invariant defined in Eq.(3) is equivalent to the invariant defined in (2).

For a system of three qubits, we can obtain

$$\begin{aligned} Z^1_{123} &= 4[(a_0 a_7 - a_1 a_6 - a_2 a_5 + a_3 a_4)^2 + 4(a_0 a_3 - a_1 a_2)(a_5 a_6 - a_4 a_7)] \\ Z^2_{123} &= 4[(a_0 a_7 - a_1 a_6 + a_2 a_5 - a_3 a_4)^2 + 4(a_0 a_5 - a_1 a_4)(a_3 a_6 - a_2 a_7)] \\ Z^3_{123} &= 4[(a_0 a_7 + a_1 a_6 - a_2 a_5 - a_3 a_4)^2 + 4(a_0 a_6 - a_2 a_4)(a_3 a_5 - a_1 a_7)] \end{aligned} \tag{4}$$

---

[*] Corresponding Author: zhxw@xiyou.edu.cn

From Eq.(4), it is direct to show that $Z_{123}^1 = Z_{123}^2 = Z_{123}^3$, thus $\tau_{123} = |Z_{123}^1|$ is just that of the 3-tangle [3].

But for the case of five qubits, we find there are five invariants

$$\begin{aligned}
Z_{12345}^1 &= \langle\psi|\hat{T}_{2y}\hat{T}_{3y}\hat{T}_{4y}\hat{T}_{5y}\hat{\sigma}_{1x}|\psi\rangle^2 + \langle\psi|\hat{T}_{2y}\hat{T}_{3y}\hat{T}_{4y}\hat{T}_{5y}\hat{\sigma}_{1z}|\psi\rangle^2 - \langle\psi|\hat{T}_{2y}\hat{T}_{3y}\hat{T}_{4y}\hat{T}_{5y}|\psi\rangle^2 \\
&= [a_0 a_{31} - a_1 a_{30} - a_2 a_{29} + a_3 a_{28} - a_4 a_{27} + a_5 a_{26} + a_6 a_{25} - a_7 a_{24} \\
&\quad - a_8 a_{23} + a_9 a_{22} + a_{10} a_{21} - a_{11} a_{20} + a_{12} a_{19} - a_{13} a_{28} - a_{14} a_{17} + a_{15} a_{16}]^2 \\
&\quad + 4(a_0 a_{15} - a_1 a_{14} - a_2 a_{13} + a_3 a_{12} - a_4 a_{11} + a_5 a_{15} + a_6 a_9 - a_7 a_8) \\
&\quad \times (-a_{16} a_{31} + a_{17} a_{30} + a_{18} a_{29} - a_{19} a_{28} + a_{20} a_{27} - a_{21} a_{26} - a_{22} a_{25} + a_{23} a_{24})
\end{aligned} \quad (5\text{-a})$$

$$\begin{aligned}
Z_{12345}^2 &= \langle\psi|\hat{T}_{1y}\hat{T}_{3y}\hat{T}_{4y}\hat{T}_{5y}\hat{\sigma}_{2x}|\psi\rangle^2 + \langle\psi|\hat{T}_{1y}\hat{T}_{3y}\hat{T}_{4y}\hat{T}_{5y}\hat{\sigma}_{2z}|\psi\rangle^2 - \langle\psi|\hat{T}_{1y}\hat{T}_{3y}\hat{T}_{4y}\hat{T}_{5y}|\psi\rangle^2 \\
&= [a_0 a_{31} - a_1 a_{30} - a_2 a_{29} + a_3 a_{28} - a_4 a_{27} + a_5 a_{26} + a_6 a_{25} - a_7 a_{24} \\
&\quad + a_8 a_{23} - a_9 a_{22} - a_{10} a_{21} + a_{11} a_{20} - a_{12} a_{19} + a_{13} a_{28} + a_{14} a_{17} - a_{15} a_{16}]^2 \\
&\quad + 4(a_0 a_{23} - a_1 a_{22} - a_2 a_{21} + a_3 a_{20} - a_4 a_{19} + a_5 a_{18} + a_6 a_{17} - a_7 a_{16}) \\
&\quad \times (-a_8 a_{31} + a_9 a_{30} + a_{10} a_{29} - a_{11} a_{28} + a_{12} a_{27} - a_{13} a_{26} - a_{14} a_{25} + a_{15} a_{24})
\end{aligned} \quad (5\text{-b})$$

$$\begin{aligned}
Z_{12345}^3 &= \langle\psi|\hat{T}_{1y}\hat{T}_{2y}\hat{T}_{4y}\hat{T}_{5y}\hat{\sigma}_{3x}|\psi\rangle^2 + \langle\psi|\hat{T}_{1y}\hat{T}_{2y}\hat{T}_{4y}\hat{T}_{5y}\hat{\sigma}_{3z}|\psi\rangle^2 - \langle\psi|\hat{T}_{1y}\hat{T}_{2y}\hat{T}_{4y}\hat{T}_{5y}|\psi\rangle^2 \\
&= [a_0 a_{31} - a_1 a_{30} - a_2 a_{29} + a_3 a_{28} + a_4 a_{27} - a_5 a_{26} - a_6 a_{25} + a_7 a_{24} \\
&\quad - a_8 a_{23} + a_9 a_{22} + a_{10} a_{21} - a_{11} a_{20} - a_{12} a_{19} + a_{13} a_{28} + a_{14} a_{17} - a_{15} a_{16}]^2 \\
&\quad + 4(a_0 a_{27} - a_1 a_{26} - a_2 a_{25} + a_3 a_{24} - a_8 a_{19} + a_9 a_{18} + a_{10} a_{17} - a_{11} a_{16}) \\
&\quad \times (-a_4 a_{31} + a_5 a_{30} + a_6 a_{29} - a_7 a_{28} + a_{12} a_{23} - a_{13} a_{22} - a_{14} a_{21} + a_{15} a_{20})
\end{aligned} \quad (5\text{-c})$$

$$\begin{aligned}
Z_{12345}^4 &= \langle\psi|\hat{T}_{1y}\hat{T}_{2y}\hat{T}_{3y}\hat{T}_{5y}\hat{\sigma}_{4x}|\psi\rangle^2 + \langle\psi|\hat{T}_{1y}\hat{T}_{2y}\hat{T}_{3y}\hat{T}_{5y}\hat{\sigma}_{4z}|\psi\rangle^2 - \langle\psi|\hat{T}_{1y}\hat{T}_{2y}\hat{T}_{3y}\hat{T}_{5y}|\psi\rangle^2 \\
&= [a_0 a_{31} - a_1 a_{30} + a_2 a_{29} - a_3 a_{28} - a_4 a_{27} + a_5 a_{26} - a_6 a_{25} + a_7 a_{24} \\
&\quad - a_8 a_{23} + a_9 a_{22} - a_{10} a_{21} + a_{11} a_{20} + a_{12} a_{19} - a_{13} a_{28} + a_{14} a_{17} - a_{15} a_{16}]^2 \\
&\quad + 4(a_0 a_{29} - a_1 a_{28} - a_4 a_{25} + a_5 a_{24} - a_8 a_{21} + a_9 a_{20} + a_{12} a_{17} - a_{13} a_{16}) \\
&\quad \times (-a_2 a_{31} + a_3 a_{30} + a_6 a_{27} - a_7 a_{26} + a_{10} a_{23} - a_{11} a_{22} - a_{14} a_{19} + a_{15} a_{18})
\end{aligned} \quad (5\text{-d})$$

$$\begin{aligned}
Z_{12345}^5 &= \langle\psi|\hat{T}_{1y}\hat{T}_{2y}\hat{T}_{3y}\hat{T}_{4y}\hat{\sigma}_{5x}|\psi\rangle^2 + \langle\psi|\hat{T}_{1y}\hat{T}_{2y}\hat{T}_{3y}\hat{T}_{4y}\hat{\sigma}_{5z}|\psi\rangle^2 - \langle\psi|\hat{T}_{1y}\hat{T}_{2y}\hat{T}_{3y}\hat{T}_{4y}|\psi\rangle^2 \\
&= [a_0 a_{31} + a_1 a_{30} - a_2 a_{29} - a_3 a_{28} - a_4 a_{27} - a_5 a_{26} + a_6 a_{25} + a_7 a_{24} \\
&\quad - a_8 a_{23} - a_9 a_{22} + a_{10} a_{21} + a_{11} a_{20} + a_{12} a_{19} + a_{13} a_{28} - a_{14} a_{17} - a_{15} a_{16}]^2 \\
&\quad + 4(a_0 a_{30} - a_2 a_{28} - a_4 a_{26} + a_6 a_{24} - a_8 a_{22} + a_{10} a_{20} + a_{12} a_{18} - a_{14} a_{16}) \\
&\quad \times (-a_1 a_{31} + a_3 a_{29} + a_5 a_{27} - a_7 a_{25} + a_9 a_{23} - a_{11} a_{21} - a_{13} a_{19} + a_{15} a_{17})
\end{aligned} \quad (5\text{-e})$$

Here $Z_{12345}^1$ is just that of $A^*$ in Ref. [1]. From the above equations, it is straightforward to prove that $Z_{12345}^i \neq Z_{12345}^j$ ($i \neq j$) for a general state. However, according to their definition, there should be five residual entanglements since any two polynomial invariants of Eq.(5) are unequal. As everyone know, this is impossible, thus the residual entanglement defined in Ref.[1] cannot extended to the case of five qubits, never to say the general odd-qubit case.

## Acknowledgements

This work is supported by Shaanxi Natural Science Foundation under Contract Nos. 2004A15 and Science Plan Foundation of office the Education Department of Shaanxi Province Contract Nos. 05JK288.

## References

1. D. Li *et al.*, Phys. Rev. A 76, 032304, (2007).
2.X W Zha and C M Zhang    e-print quant-ph/0702046.
3. V. Coffman, J. Kundu, and W. K. Wootters, Phys. Rev. A **61**, 052306 (2000).